\documentclass[a4paper,reqno,10pt,english]{amsart} 
\usepackage[cp1252]{inputenc}
\usepackage{graphicx}
\usepackage[all]{xy}
\usepackage{enumerate}
\usepackage{array}
\usepackage{amssymb}
\usepackage{amscd}
\usepackage{color}
\usepackage{hyperref}
\theoremstyle{plain}
\newtheorem{prop}{Proposition}[section]
\newtheorem{teo}[prop]{Theorem}
\newtheorem{cor}[prop]{Corollary}
\newtheorem{lema}[prop]{Lemma}
\newtheorem{defn}[prop]{Definition}
\newtheorem{obs}[prop]{Remark}
\theoremstyle{definition}
\newtheorem{ejem}{Example}

\theoremstyle{remark}

\numberwithin{equation}{section}

\newcommand{\St}{St(x,\Gamma_{n})}
\newcommand{\ef}{\mathbf{\Gamma}} 

\begin{document}

\title[Fractal Dimension III: A Hausdorff Approach]{Fractal Dimension for Fractal Structures: A Hausdorff Approach}
\author{M. Fern\'andez-Mart\'{\i}nez and M.A S\'anchez-Granero}
\curraddr{Area of Geometry and Topology \\ Faculty of
Science \\
Universidad de Almer\'{\i}a \\ 04071 Almer\'{\i}a \\ Spain}
\email{\url{fmm124@ual.es} \and \url{misanche@ual.es}}
\thanks{The second author acknowledges the support of the Spanish Ministry
of Science and Innovation, grant MTM2009-12872-C02-01.}



\begin{abstract}
This paper provides a new model
to compute the fractal dimension of a subset on a generalized-fractal space.
Recall that fractal structures are a perfect place where a new definition of fractal dimension can be given,
so we
perform a suitable discretization
of the Hausdorff theory of fractal dimension.
We also find some connections between our definition and the classical ones and also with fractal dimensions I \& II (see \cite{SGFD}).
Therefore, we generalize them and obtain an easy method in order to calculate the fractal dimension of strict self-similar sets which are not required to verify the \emph{open set condition}.

\end{abstract}

\maketitle
\parskip 2ex

\section{Introduction}
The analysis and study of fractals has become very important during last years since the large number of applications to diverse fields that these kind of sets have experimented.
In this way, the introduction of fractal structures has allowed to formalize some topics on fractal theory from both theoretical and applied points of view.
Furthermore, the use of fractal structures leads to connect diverse interesting topics on topology like transitive quasi-uniformities, non-archimedean quasimetrization, metrization, topological and fractal dimensions, self-similar sets and even space filling curves (see \cite{SG10}).
Indeed, one of the main tools applied to the study of fractals is the fractal dimension, understood as the classical box-counting and Hausdorff dimension, since it is a single quantity which offers some information about the complexity of the set under study.
These two notions can be defined for any metrizable space, and while the former is \emph{better} from an applied point of view, the latter presents \emph{better} analytical properties, though it can result very difficult or impossible to calculate in practical applications.

Likewise, fractal dimension theory has been applied in some fields of
science, such as the study of dynamical systems (\cite{FAN10}), diagnosis of diseases, such as osteoporosis (\cite{RUT92}) or cancer (\cite{BAI00}),
ecology (\cite{BAR03}), earthquakes (\cite{HIR89}), detection of
eyes in human face images (\cite{LIN01}), and
the analysis of the human retina (\cite{LAN03}), just to name a few.

One of the main goal of this paper consists of providing a new model in order to compute the fractal dimension of a given subset on the context of fractal structures (by means of GF-spaces), which has interesting theoretical properties, but could also be calculated with easiness.
In this way, we consider a suitable discretization of the underlying idea on the construction of Hausdorff measure and dimension, and propose a new definition of fractal dimension on a generalized-fractal space.

First of all, we motivate the way it is defined and then we obtain an easy method in order to calculate it from an effective point of view.
We also find some relations between the classical fractal dimension definitions and the so called fractal dimension III, and in particular, we obtain some interesting properties over the elements of a fractal structure in order to get the equality with those definitions of fractal dimension defined on a GF-space (see \cite{SGFD}). Thus, the new definition generalizes them as well as box-counting dimension, which can be obtained as a particular case.
On the other way, self-similar sets provide a particular kind of fractals which have a fractal structure on a natural way, which allows to study them since the point of view of GF-spaces. Taking it into account, we show that its structure as well as the definition of fractal dimension III, leads to obtain the fractal dimension of a strict self-similar set by solving an easy equation which only involves the similarity factors associated with the corresponding IFS.
With this in mind, we have that this result does not require to verify the open set condition hypothesis used in the classical theorems.

\section{Preliminaries}
We start with some preliminary topics.

\subsection{Fractal structures and self-similar sets}\label{sec:1}
The main purpose of this section consists of recalling some notations and basic notions that will be useful in this paper.

In this way, the key concept we are going to use is about \emph{fractal structures}. Nevertheless, although a more natural use of them is in the study of fractals, and in particular self-similar sets (see \cite{SGnd}), its introduction was first motivated in order to characterize non-archimedeanly quasimetrization (see \cite{SG99}).
The use of fractal structures provides a powerful tool in order to study new models for a fractal dimension definition, since they will allow to distinguish and classify a larger volume of spaces than by using the classical definitions of fractal dimension (which will be obtained as a particular case), that only work over metrizable spaces.
So that, these kind of topological spaces constitutes a perfect place in order to develop a theory on fractal dimension.

Let $\Gamma$ be a covering of $X$. Thus, we will denote $St(x,\Gamma)=\cup\{A\in
\Gamma:x\in A\}$ and $U_{x \Gamma}=X \setminus \cup \{A\in
\Gamma:x\notin A\}$.
Furthermore, if $\ef=\{\Gamma_{n}:n\in \mathbb{N}\}$ is a countable family of coverings of
$X$, then we will denote $U_{xn}=U_{x
\Gamma_n}$, $\mathcal{U}_{x}^{\ef}=\{U_{xn}:n\in \mathbb{N}\}$
and $St(x,\ef)=\{\St:n\in \mathbb{N}\}$ .

The next definition was introduced in \cite{SG99}.
\begin{defn}
Let $X$ be a topological space. A pre-fractal structure on $X$ is a countable
family of coverings (called levels) $\ef=\{\Gamma_{n}:n\in \mathbb{N}\}$ such
that $\mathcal{U}_{x}^{\ef}$  is an open neighborhood
base of $x$ for each $x\in X$.\newline
Moreover, if $\Gamma_{n+1}$ is a refinement of $\Gamma_{n}$ (which can be denoted by $\Gamma_{n+1}\prec \Gamma_{n}$), such that for all $x\in A$ with $A\in \Gamma_{n}$, there exists $B\in \Gamma_{n+1}$ such that $x\in B\subseteq A$, we will say that $\ef$ is a fractal structure on $X$.\newline
If $\ef$ is a (pre-) fractal structure on $X$, then we will say that $(X,\ef)$
is a generalized (pre-) fractal space, or simply a (pre-) GF-space. If there is
no doubt about $\ef$, then we will say that $X$ is a (pre-) GF-space.
\end{defn}

\begin{obs}
In this paper, the levels we use in order to define a fractal structure $\ef$ are not coverings in the usual sense, since we are going to enable the possibility that could exist elements on any level of the fractal structure which can appear twice or more times in the same level. For instance, $\Gamma_{1}=\{[0,\frac{1}{2}],[\frac{1}{2},1],[0,\frac{1}{2}]\}$ could be the first level of a given fractal structure $\ef$ defined over the closed unit interval.
\end{obs}

Note also that if $\ef$ is a pre-fractal structure, then any of its levels is a closure preserving closed covering (see [\cite{SG02B},Prop. 2.4]).

If $\ef$ is a fractal structure on $X$ and $St(x,\ef)$ is a neighborhood base of $x$ for all $x\in X$, we will call $\ef$ a \emph{starbase} fractal structure.
In general, if $\Gamma_{n}$ has the property $P$ for all $n\in \mathbb{N}$, and
$\ef=\{\Gamma_{n}:n\in \mathbb{N}\}$ is a fractal structure on $X$, we will say that $\ef$ is a fractal
structure with the property $P$, and that $(X,\ef)$ is a GF-space with
the property $P$.
For instance, if $\Gamma_{n}$ is a finite covering for all natural number $n$ and $\ef$ is a fractal structure on $X$, then we will say that $\ef$ is a finite fractal structure on $X$, and that $(X,\ef)$ is a finite GF-space.

On the other hand, we also recall the definition of self-similar set provided by Hutchinson (see \cite{HUT81}).

\begin{defn}
Let $I=\{1,\ldots, m\}$ be a finite index set and let $\{f_{i}:i\in I\}$ be a family of contractive mappings defined from a complete metric space $X$ into itself.
Then there exists a unique non-empty compact subset $K$ of $X$ such that $K=\cup_{i\in I}f_{i}(K)$, which is called a self-similar set.
\end{defn}
In classical non-linear theory, $(X,\{f_{i}:i\in I\})$ is called an iterated function scheme (which we will denote by IFS for short), and the self-similar set $K$, is the \emph{atractor} of that IFS.
Next, we provide an interesting example which describes analytically the so called Sierpinski's gasket, which is a typical example of a strict self-similar set.

\begin{ejem}\label{ejem:1}
Let $I=\{1,2,3\}$ be a finite index set and let $\{f_{i}:i\in I\}$ be a finite set of similarities over the euclidean plane which are defined by
\begin{equation}\label{eq:48}
f_{i}(x,y)=\left\{
\begin{array}{ll}
\hbox{$(\frac{x}{2},\frac{y}{2})$}\ \hfill \text{if} & \hbox{$i=1$} \\
\hbox{$f_{1}(x,y)+(\frac{1}{2},0)$}\ \hfill\text{if} & \hbox{$i=2$} \\
\hbox{$f_{1}(x,y)+(\frac{1}{4},\frac{1}{2})$}\ \hfill\text{if} & \hbox{$i=3$}
\end{array}
\right.
\end{equation}
for all $(x,y)\in \mathbb{R}^{2}$. Thus, the Sierpinski's gasket is determined on an unique way as the non-empty compact subset verifying the next Hutchinson's equation:
$K=\cup_{i\in I}f_{i}(K)$. In this way,
note that each component $f_{i}(K)$ is a self-similar copy of the atractor of the IFS $(\mathbb{R}^{2},\{f_{i}:i\in I\})$.
\end{ejem}
Self-similar sets constitute an interesting kind of fractals that are
characterized by having a
fractal structure in a natural way, which was first sketched in \cite{BA92}. Indeed, that paper becomes the origin of the term \emph{fractal structure}.
Next, we present the description of such fractal structure
(see \cite{SGnd}).

\begin{defn}
Let $I=\{1,\ldots, m\}$ be a finite
index set, and let $(X,\{f_{i}:i\in I\})$ be an IFS whose associated
self-similar set is $K$. The natural fractal structure on
$K$ can be defined as the countable family of coverings $\ef=\{\Gamma_{n}:n\in
\mathbb{N}\}$, where $\Gamma_{n}=\{f_{\omega}(K):\omega\in I^{n}\}$ for every
natural number $n$. Here for all $n\in \mathbb{N}$ and all
$\omega=\omega_{1} \ \omega_{2}\ \ldots \ \omega_{n}\in I^{n}$, we denote
$f_{\omega}^{n}=f_{\omega_{1}}\ \circ \ \ldots \ \circ \ f_{\omega_{n}}$.

This fractal structure can also be described as follows:
$\Gamma_{1}=\{f_{i}(K):i\in I\}$ and $\Gamma_{n+1}=\{f_{i}(A): A\in \Gamma_{n},
i\in I\}$ for all $n\in \mathbb{N}$.
\end{defn}

On example \ref{ejem:1} we described analytically the IFS whose associated
self-similar set is the Sierpinski's triangle. Next, we are going to present
the natural fractal structure associated with this strict self-similar set.
\begin{ejem}
The natural fractal structure associated with the Sierpinski's
triangle can be described as the countable family of coverings
$\ef=\{\Gamma_{n}: n\in \mathbb{N}\}$, where $\Gamma_{1}$ is the union of three
equilateral ``triangles'' with sides equal to $\frac{1}{2}$, $\Gamma_{2}$
consists of the union of $3^{2}$ equilateral ``triangles'' with sides equal to
$\frac{1}{2^{2}}$, and in general, $\Gamma_{n}$ is the union of $3^{n}$
equilateral ``triangles'' whose sides are equal to $\frac{1}{2^{n}}$.
Furthermore, this is a finite starbase 
fractal structure.
\end{ejem}


\subsection{Box-counting dimension and fractal dimensions I \& II}
Fractal dimension is one of the main tools used in order to study fractals, since it is a single value which provides information about its complexity and the irregularities it presents when being examined with enough level of detail. In this way, the fractal dimension is understood as the classical box-counting and Hausdorff dimensions. Note that both of them can be defined over any metrizable space. Thus, while the former is \emph{better} from an applied point of view, the latter is \emph{better} from a theoretical point of view. The basic theory on Hausdorff and box-counting dimensions can be found in \cite{FAL90}.\newline
One of the main advantages while working with box-counting dimension consists of its effective calculation an empirical estimation. This fractal dimension has been also known as \emph{information dimension, Kolmogorov entropy, capacity dimension, entropy dimension, metric dimension , \ldots} etc.
Thus, the (lower/upper) box-counting dimensions of a subset $F\subset \mathbb{R}^{d}$ are given by the following (lower/upper) limit:
\begin{equation}\label{eq:81}
\dim_{B}(F)=\lim_{\delta\rightarrow 0}\frac{\log N_{\delta}(F)}{-\log \delta}
\end{equation}
where $\delta$ is the scale used in the study of $F$ and $N_{\delta}(F)$ is the number of $\delta-$cubes which meet $F$. Recall that a $\delta-$cube in $\mathbb{R}^{d}$ is a set of the form $[k_{1}\ \delta,(k_{1}+1)\ \delta]\times \ldots \times [k_{d}\ \delta,(k_{d}+1)\ \delta]$ where $k_{i}$ are integers for all $i\in \{1,\ldots, d\}$. Note that the limit given at \ref{eq:81} can be discretized by means of $\delta=\frac{1}{2^{n}}$.
Therefore, the box-counting dimension can be estimated as the slope of a log-log graph plotted over a suitable discrete collection of scales $\delta$.\newline
Recall that the natural fractal structure on the euclidean space $\mathbb{R}^{d}$ is defined by $\ef=\{\Gamma_{n}:n\in \mathbb{N}\}$, where the levels are given by $\Gamma_{n}=\{[\frac{k_{1}}{2^{n}},\frac{k_{1}+1}{2^{n}}]\times [\frac{k_{2}}{2^{n}},\frac{k_{2}+1}{2^{n}}]\times \ldots \times [\frac{k_{d}}{2^{n}},\frac{k_{d}+1}{2^{n}}]:k_{i}\in \mathbb{Z}, i\in \{1,\ldots, d\}\}$ for all $n\in \mathbb{N}$. In this way, if we select $\delta=\frac{1}{2^{n}}$, then $N_{\delta}(F)$ is just the number of elements of each level $\Gamma_{n}$ of the fractal structure which meet $F$.
Hence a natural idea arises: we can propose a new definition of fractal dimension for any fractal structure which generalizes the classical box-counting dimension. Indeed, it agrees with the latter when taking $\ef$ as the natural fractal structure on $\mathbb{R}^{d}$. Note that the definition of a fractal dimension on a GF-space allows to classify and distinguish a larger volume of spaces than by means of the classical box-counting dimension definition. It results also useful in order to calculate the fractal dimension over another kind of spaces, such as the non-euclidean ones, where the box-counting dimension can have no sense or can be difficult or impossible to calculate.

Accordingly, we recall the next two models in order to determine the fractal dimension of a subset $F$ of a GF-space $(X,\ef)$ introduced in \cite{SGFD}. The (lower/upper) fractal dimensions I \& II are defined by the following (lower/upper) limits:
\begin{equation}\label{eq:82}
\dim_{\ef}^{1}(F)=\lim_{n\rightarrow \infty}\frac{\log N_{n}(F)}{n\log 2}
\end{equation}
\begin{equation}\label{eq:83}
\dim_{\ef}^{2}(F)=\lim_{n\rightarrow \infty}\frac{\log N_{n}(F)}{-\log \delta(F,\Gamma_{n})}
\end{equation}
where $N_{n}(F)$ is the number of elements of the level $\Gamma_{n}$ of the fractal structure which meet $F$, and $\delta(F,\Gamma_{n})$ is the supremum of the diameters of all the elements of each level $\Gamma_{n}$ which meet $F$.
Note that the fractal dimension II given at \ref{eq:83} is calculated respect to a fractal structure $\ef$ associated to a metric (or a distance) space $(X,\rho)$, while the fractal dimension I does not depend on any metric.


\section{A new model based on the Hausdorff scheme in order to determine the fractal dimension on GF-Spaces}
\subsection{Introduction \& Motivation}

In \cite{SGFD}, we investigated two different ways from both theoretical and applied viewpoints, in order to calculate the fractal dimension of any subset of a generalized-fractal space. In this way, recall that fractal structures are a nice place where a new model to determine the fractal dimension of a subset can be given.
Recall that the fractal dimension I formula allows the use of a larger collection of fractal structures than the box-counting expression in order to compute it: in particular, the natural fractal structure on any euclidean space.
On the other hand, although the second model we considered enables the possibility that could exist different diameter sets on each level of the fractal structure, it does not distinguish between different diameter sets (see \cite[Remark 4.6]{SGFD}).
Note that we have to count the number of elements of any level of the fractal structure which meet the subset whose fractal dimension we want to calculate. Then, we weigh it by means of a discrete scale: a fixed quantity on each level (fractal dimension I), or the largest diameter of the elements on this level (fractal dimension II). This idea is inspired on a suitable \emph{discretization} of the box-counting dimension definition.

Hausdorff dimension provides another interesting philosophy in order to calculate the fractal dimension of a given subset of a metrizable space. Indeed, let $(X,d)$ be a metric space. Given a scale $\delta>0$ and a subset $F$ of $X$, recall that a $\delta-$cover is a countable family of subsets $\{U_{i}\}_{i\in I}$ such that $F\subseteq \cup_{i\in I}U_{i}$, with $\delta(U_{i})\leq \delta$ for all $i\in I$. The underlying idea on Hausdorff dimension is based on the Hausdorff measure, and it consists of minimizing the sum of the $s-$powers of all the diameters of the subsets on any $\delta-$cover, where $s$ is going to be the fractal dimension we are looking for. In this way, the following quantity is defined:
\begin{equation}
\mathcal{H}_{\delta}^{s}(F)=\inf\Bigg\{\sum_{i=1}^{\infty}\delta(U_{i})^{s}: \{U_{i}\}_{i\in I} \ \text{is a} \ \delta- \ \text{cover of} \ F\Bigg\}
\end{equation}
Note that when $\delta$ decreases, then the class of $\delta-$coverings is reduced, so that the measure of $F$ increases. Thus, the next limit always exists:
\begin{equation*}
\mathcal{H}_{H}^{s}(F)=\lim_{\delta\rightarrow 0}\mathcal{H}_{\delta}^{s}(F)
\end{equation*}
which is called the $s-$dimensional Hausdorff measure of $F$.
Then, the Hausdorff dimension is characterized as the point where $\mathcal{H}_{H}^{s}(F)$ leaves the quantity $\infty$ and reaches the value $0$, namely:
\begin{equation}\label{eq:71}
\dim_{H}(F)=\inf\{s:\mathcal{H}_{H}^{s}(F)=0\}=\sup\{s:\mathcal{H}_{H}^{s}(F)=\infty\}
\end{equation}
Equivalently, we have that follows:
\begin{equation*}
\mathcal{H}_{H}^{s}(F)=\left\{
\begin{array}{ll}
\hbox{$\infty$}\ \ldots & \hbox{$s<\dim_{H}(F)$} \\
\hbox{$0$}\ \ldots & \hbox{$s>\dim_{H}(F)$}
\end{array}
\right.
\end{equation*}
In particular, if $s=\dim_{H}(F)$, then $\mathcal{H}_{H}^{s}(F)$ can be equal to $0$, $\infty$, and even it can be possible that $\mathcal{H}_{H}^{s}(F)\in (0,\infty)$.

Recently, Urba\'nski (\cite{UR10}) defined a kind of transfinite version of the
Hausdorff dimension for a deeper study of spaces with infinite Hausdorff
dimension.

On the other hand, our main purpose consists of providing a new definition of
fractal dimension on the more general context of GF-spaces, by means of the
underlying ideas of Hausdorff dimension on any metrizable space.
In this way, let $\ef$ be a fractal structure on a metric space $(X,\rho)$, and let $F$ be a subset of $X$. The main idea in order to get a more accurate value for the fractal dimension of $F$ is going to take into account the size of all the elements on any level of the family $\ef$ which meet $F$, by means of its diameters. Thus, consider the next family of elements of $\ef$:
\begin{equation}\label{eq:80}
\mathcal{A}_{n}(F)=\{A\in \Gamma_{n}:A\cap F\neq \emptyset\}
\end{equation}
for all $n\in \mathbb{N}$, and let $s$ be a positive real number.
Hence, the sum of the $s-$powers of the diameters of all the elements of each family $\mathcal{A}_{n}(F)$ could determine us a register of how irregular the set $F$ is, providing an approximation of its evolution and complexity. Equivalently, we are going to begin with an expression like the following:
\begin{equation}\label{eq:63}
\mathcal{H}_{n}^{s}(F)=\sum\{\delta(A)^{s}:A\in \mathcal{A}_{n}(F)\}
\end{equation}
for all natural number $n$.
By means of the previous discrete sequence, define also the following value:
\begin{equation*}\label{eq:64}
\mathcal{H}^{s}(F)=\lim_{n\rightarrow \infty}\mathcal{H}_{n}^{s}(F)
\end{equation*}
In this way, as well as Hausdorff dimension verifies an expression like \ref{eq:71}, we are going to explore a property of this kind for the new model by taking into account the expressions \ref{eq:64} and \ref{eq:63}.
Indeed, let $t$ be another positive real number. Then, note that
\begin{equation*}\label{eq:67}
\sum \delta(A)^{t}\leq \delta_{n}^{t-s}\sum \delta(A)^{s}
\end{equation*}
where the sums are considered over any family $\mathcal{A}_{n}(F)$, and $\delta_{n}=\delta(F,\Gamma_{n})$ for all $n\in \mathbb{N}$.  The previous inequality is equivalent to $\mathcal{H}_{n}^{t}(F)\leq \delta_{n}^{t-s} \ \mathcal{H}_{n}^{s}(F)$ for all natural number $n$.
In this way, taking limits as $n\rightarrow \infty$ on the previous expression leads to
\begin{equation*}
\mathcal{H}^{t}(F)\leq \mathcal{H}^{s}(F)\cdot \lim_{n\rightarrow \infty}\delta_{n}^{t-s}
\end{equation*}

if $\mathcal{H}^{s}(F)<\infty$.
Furthermore, if $\mathcal{H}^{s}(F)<\infty$ and $\delta_{n}\rightarrow 0$ as $n\rightarrow \infty$ with $t>s$, then we have that $\mathcal{H}^{t}(F)=0$.
Accordingly, under the natural hypothesis that $\delta(F,\Gamma_{n})$ converges to $0$, the new theoretical method in order to calculate the fractal dimension on a GF-space determines that this value is the point where $\mathcal{H}^{s}(F)$ \emph{jumps} from $\infty$ to $0$. Equivalently, if we denote by $\dim_{\ef}$ to the new fractal dimension, we have again that
\begin{equation*}
\dim_{\ef}(F)=\inf\{s:\mathcal{H}^{s}(F)=0\}=\sup\{s:\mathcal{H}^{s}(F)=\infty\}
\end{equation*}
whenever $\delta_{n}\rightarrow 0$. Furthermore, this hypothesis is going to be necessary, as the next remark shows, and note that it is only a natural restriction over the size of the elements on each level of the fractal structure.

\begin{obs}\label{obs:4}
There exists a fractal structure $\ef$ on a metric space $(X,\rho)$ and a subset $F$ of $X$ such that $\delta(F,\Gamma_{n})\nrightarrow 0$ verifying that $\inf\{s:\mathcal{H}^{s}(F)=0\}\neq \sup\{s:\mathcal{H}^{s}(F)=\infty\}$.
\end{obs}

\begin{proof}
Let  $F=[0,1]\times [0,1]$ and let $\ef$ be the natural fractal structure on the euclidean plane induced on the unit square but add $F$ itself to all levels of the fractal structure.
Thus, if we take into account \ref{eq:63} in order to determine the fractal dimension of $F$, then we can check that the plot which compares $\mathcal{H}^{s}(F)$ versus $s$ does not present a behavior like the Hausdorff model for fractal dimension.
Indeed, note that $\delta(F,\Gamma_{n})=\sqrt{2}$ for all $n\in \mathbb{N}$, which implies that $\delta(F,\Gamma_{n})\nrightarrow 0$ as $n\rightarrow \infty$.
On the other hand, we obtain that $\mathcal{H}_{n}^{s}(F)=2^{\frac{s}{2}}\cdot \Big(1+\frac{1}{2^{n(s-2)}}\Big)$ for all natural number $n$.
Hence, it is clear that
\begin{equation*}
\mathcal{H}^{s}(F)=\left\{
\begin{array}{ll}
\hbox{$2^{\frac{s}{2}}$}\ \hfill \ldots & \hbox{$s>2$} \\
\hbox{$\infty$}\ \hfill \ldots & \hbox{$s<2$}
\end{array}
\right.
\end{equation*}
Accordingly, we have that $\sup\{s:\mathcal{H}^{s}(F)=\infty\}\neq\inf\{s:\mathcal{H}^{s}(F)=0\}$.
\end{proof}

\begin{center}
\begin{figure*}[here]
\begin{tabular}{c}
\includegraphics[width=115mm, height=55mm]{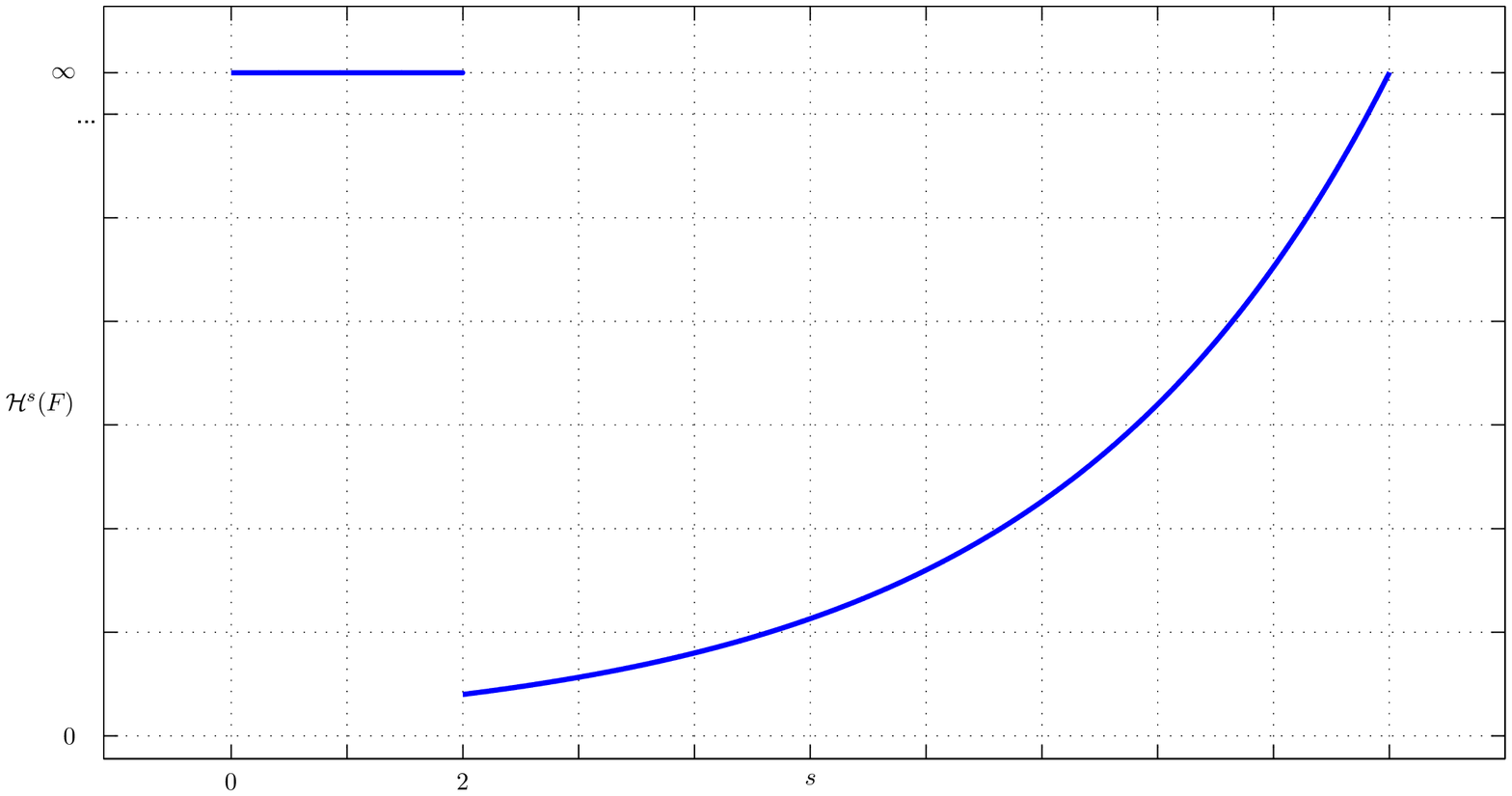} \\ \includegraphics[width=115mm, height=55mm]{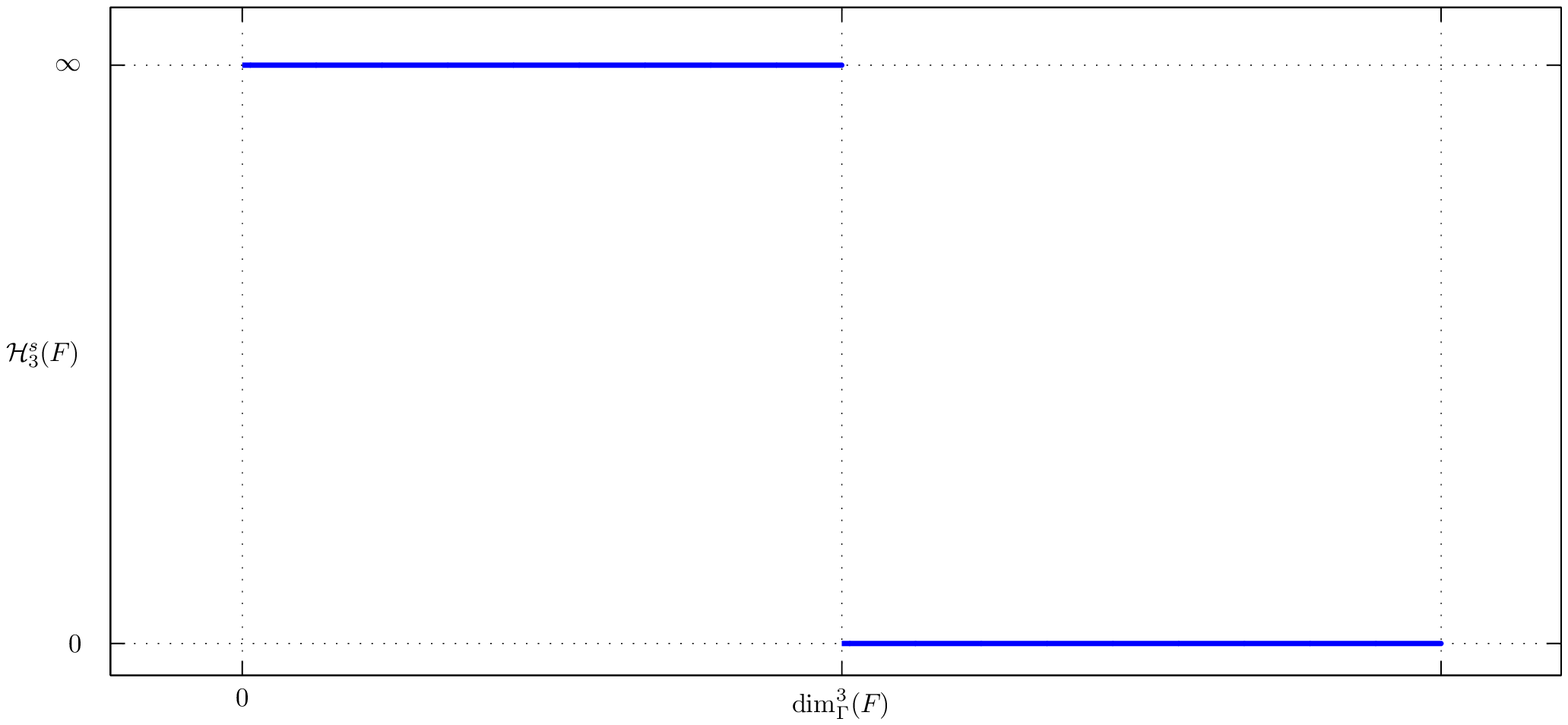}
\end{tabular}
\caption{Graph representation of $\mathcal{H}^{s}(F)$ versus $s$ for a set $F$ such that $\delta(F,\Gamma_{n})\nrightarrow 0$ (see remark \ref{obs:4}), and plot of $\mathcal{H}_{3}^{s}(F)$ versus $s$ by means of definition \ref{eq:69} of fractal dimension (bottom).}
\end{figure*}
\end{center}


%
%

Nevertheless, unlike it happens with $\mathcal{H}_{H}^{s}$ (which always exists for all subset of $X$), we have that the set function $\mathcal{H}_{n}^{s}$ given at \ref{eq:63} is not a monotonic sequence on $n\in \mathbb{N}$, which implies that $\mathcal{H}^{s}(F)$ does not always exist in general. This fact leads to
define lower/upper $\mathcal{H}^{s}(F)$, by means of lower/upper versions of the next limit:
\begin{equation}
\mathcal{H}^{s}(F)=\lim_{n\rightarrow \infty}\mathcal{H}_{n}^{s}(F)
\end{equation}

In order to solve the problem consisting of the existence of the set function $\mathcal{H}^{s}$, we are going to take into account the following family of elements of the fractal structure, instead of the previous $\mathcal{A}_{n}(F)$:
\begin{equation}\label{eq:66}
\mathcal{A}_{n,3}(F)=\cup_{m\geq n}\{\mathcal{A}_{m}(F)\}
\end{equation}
for all natural number $n$. Using each family $\mathcal{A}_{n,3}(F)$ given at \ref{eq:66}, the previous arguments remains valid.


\subsection{The key concept \& Results}
With this in mind, next we are going to provide the key definition of fractal dimension we are going to analyze in this paper.
\begin{defn}\label{eq:69}
Let $\ef$ be a fractal structure on a metric space $(X,\rho)$, and let $F$ be a subset of $X$. Suppose that $\delta(F,\Gamma_{n})\rightarrow 0$ and consider the following expression:
\begin{equation}\label{eq:68}
\mathcal{H}_{n,3}^{s}(F)
=\inf_{m\geq n}\mathcal{H}_{m}^{s}(F)
\end{equation}
Take also $\mathcal{H}_{3}^{s}(F)=\lim_{n\rightarrow \infty}\mathcal{H}_{n,3}^{s}(F)$. Then, the fractal dimension III of $F$ is defined by
\begin{equation*}
\dim_{\ef}^{3}(F)=\inf\{s: \mathcal{H}_{3}^{s}(F)=0\}=\sup\{s:\mathcal{H}_{3}^{s}(F)=\infty\}
\end{equation*}
\end{defn}

\begin{obs}\label{eq:86}
Note that the following items are equivalent in order to calculate $\mathcal{H}_{n,3}^{s}(F)$ for all subset $F$ of $X$ and all natural number $n$ (see \ref{eq:68}):
\begin{enumerate}
\item $\inf\{\mathcal{H}_{m}^{s}(F):m\geq n\}$.
\item $\inf\Big\{\sum_{A\in \mathcal{B}} \delta(A)^{s}: \mathcal{B}\in \mathcal{A}_{n,3}(F)\Big\}$.
\item $\inf\Big\{\sum_{A\in \mathcal{A}_{m}(K)} \delta(A)^{s}: m\geq n\Big\}$.
\end{enumerate}
\end{obs}
Thus, as happens with both Hausdorff measure and dimension, and the set function $\mathcal{H}_{H}^{s}$, we have that
\begin{equation*}
\mathcal{H}_{3}^{s}(F)=\left\{
\begin{array}{ll}
\hbox{$\infty$}\ \ldots & \hbox{$s<\dim_{\ef}^{3}(F)$} \\
\hbox{$0$}\ \ldots & \hbox{$s>\dim_{\ef}^{3}(F)$}
\end{array}
\right.
\end{equation*}
whenever $\delta(F,\Gamma_{n})\rightarrow 0$. The next remark results useful, since it asserts that we do not need to consider lower/upper limits when defining $\mathcal{H}_{3}^{s}(F)$.

\begin{obs}\label{obs:9}
Since $\mathcal{H}_{n,3}^{s}(F)$ is a monotonic sequence on $n\in \mathbb{N}$, we have that the fractal dimension III of any subset $F$ of $X$ always exists.
\end{obs}

The next theorem shows some relations between the definition \ref{eq:69} of fractal dimension and the classical definitions, namely, the box-counting and the Hausdorff dimensions.

\begin{teo}\label{teo:11}
Let $\ef$ be a fractal structure on a metric space $(X,\rho)$ and let $F$ be a subset of $X$. Suppose that $\delta(F,\Gamma_{n})\rightarrow 0$. Then,
\begin{enumerate}
\item $\dim_{\ef}^{3}(F)\leq \underline{\dim}_{\ef}^{2}(F)\leq \overline{\dim}_{\ef}^{2}(F)$.
\item If $\delta(A)=\delta(F,\Gamma_{n})$ for all $A\in \mathcal{A}_{n}(F)$, then $\underline{\dim}_{B}(F)\leq \dim_{\ef}^{3}(F)$.
\item $\dim_{H}(F)\leq \dim_{\ef}^{3}(F)$.
\end{enumerate}
\end{teo}

\begin{proof}
Let $\delta_{n}=\delta(F,\Gamma_{n})$ for each $n\in \mathbb{N}$.
\begin{enumerate}
\item Indeed, by means of expression \ref{eq:68} we have that $\mathcal{H}_{n,3}^{s}(F)\leq N_{m}(F)\cdot \delta_{m}^{s}$ for all natural number $m\geq n$.
Therefore, if $\mathcal{H}^{s}_{3}(F)>1$, then there exists $n\in \mathbb{N}$ such that $\log N_{m}(F)+s\ \log \delta_{m}\geq 0$ for all $m\geq n$.
Accordingly, we get that $\log N_{m}(F)\geq -s\ \log \delta_{m}$ for all $m\geq n$, which implies that
$s\leq \underline{\lim}_{m\rightarrow \infty}\frac{\log N_{m}(F)}{-\log \delta_{m}}$. Hence, $\dim_{\ef}^{3}(F)\leq \underline{\dim}_{\ef}^{2}(F)\leq \overline{\dim}_{\ef}^{2}(F)$.

\item First of all, in order to consider the (lower) box-counting dimension of $F$, let $N_{\delta}(F)$ be the smallest number of sets of diameter at most $\delta$ that cover $F$ (see \cite[Definition 3.1]{FAL90}).
Note that
\begin{equation*}
\underline{\dim}_{B}(F)
=\varliminf_{\delta\rightarrow 0}\frac{\log N_{\delta}(F)}{-\log \delta}
\leq \varliminf_{n\rightarrow \infty}\frac{\log N_{\delta_{n}}(F)}{-\log \delta_{n}}
\end{equation*}
Suppose that $s<\underline{\dim}_{B}(F)$. Thus, if $\varepsilon>0$ verifies that $s+\varepsilon<\varliminf_{n\rightarrow \infty}\frac{\log N_{\delta_{n}}(F)}{-\log \delta_{n}}$, then there exists $n_{0}\in \mathbb{N}$ such that $\frac{\log N_{\delta_{n}}(F)}{-\log \delta_{n}}>s+\varepsilon$ for all $n\geq n_{0}$.
Hence, $N_{\delta_{n}}(F)\cdot \delta_{n}^{s}>\delta_{n}^{-\varepsilon}$ for all $n\geq n_{0}$. Thus, $N_{\delta_{n}}(F)\cdot \delta_{n}^{s}\rightarrow \infty$.
Furthermore, note that the hypothesis consisting of $\delta(A)=\delta_{n}$ for all $A\in \mathcal{A}_{n}(F)$ leads to $N_{\delta_{n}}(F)\cdot \delta_{n}^{s}\leq \mathcal{H}_{n}^{s}(F)$ for all natural number $n$ and all $s>0$.
Accordingly, for all $R>0$ there exists $n_{0}\in \mathbb{N}$ such that it is verified that $\mathcal{H}_{m}^{s}(F)\geq N_{\delta_{m}}(F)\cdot \delta_{m}^{s}>R$ for all $m\geq n_{0}$. In particular, since $\mathcal{H}_{n_{0},3}^{s}(F)=\inf\{\mathcal{H}_{m}^{s}(F):m\geq n_{0}\}$, we obtain that $\mathcal{H}_{n_{0},3}^{s}(F)\geq R$.
Therefore, $\mathcal{H}_{n,3}^{s}(F)\geq \mathcal{H}_{n_{0},3}^{s}(F)\geq R$ for all $n\geq n_{0}$, taking into account the fact that the set function $\mathcal{H}_{n,3}^{s}$ is monotonic (see remark \ref{obs:9}). It implies that $\mathcal{H}_{3}^{s}(F)=\infty$, so that $s\leq \dim_{\ef}^{3}(F)$ for all $s<\underline{\dim}_{B}(F)$, which leads to $\underline{\dim}_{B}(F)\leq \dim_{\ef}^{3}(F)$.

\item It suffices with taking into account the fact that any covering on $\mathcal{A}_{n,3}(F)$ is also a $\delta-$covering for a suitable $\delta>0$.
\end{enumerate}
\end{proof}

Note that for a given subset $F$ of $X$, the calculation of each term of the sequence $\mathcal{H}_{n}^{s}(F)$ given by \ref{eq:63} looks easier to determine than the corresponding one on \ref{eq:68} for $\mathcal{H}_{n,3}^{s}(F)$. Nevertheless, as we have seen on definition \ref{eq:69}, the use of the set function $\mathcal{H}^{s}_{n,3}$ presents the advantage that the fractal dimension III always exists.
In this way, the next result we present is about the possibility of calculating the fractal dimension III from the expression \ref{eq:63}.

\begin{teo}\label{teo:4}
Let $\ef$ be a fractal structure on a metric space $(X,\rho)$ and let $F$ be a subset of $X$. Suppose that there exists the quantity $\mathcal{H}^{s}(F)$.
Then, the fractal dimension III of $F$ can be calculated as follows:
\begin{equation*}\label{eq:37}
\dim_{\ef}^{3}(F)=\inf\{s: \mathcal{H}^{s}(F)=0\}=\sup\{s:\mathcal{H}^{s}(F)=\infty\}
\end{equation*}
\end{teo}

\begin{proof}
First of all by means of remark \ref{eq:86}, we have that
\begin{equation*}
\mathcal{H}_{n,3}^{s}(F)=\inf\{\mathcal{H}_{m}^{s}(F):m\geq n\}
\end{equation*}
which leads to
$\mathcal{H}_{n,3}^{s}(F)\leq \mathcal{H}_{n}^{s}(F)$
for all natural number $n$ and all subset $F\subset X$.
On the other hand, let $\varepsilon$ be a fixed but arbitrary positive real number. Then, there exists $n_{0}\in \mathbb{N}$ such that
\begin{equation}\label{eq:34}
|\mathcal{H}^{s}(F)-\mathcal{H}_{n}^{s}(F)|\leq \varepsilon
\end{equation}
for all $n\geq n_{0}$. Accordingly, we have that $\mathcal{H}^{s}(F)\leq \mathcal{H}_{n}^{s}(F)+\varepsilon$ for all $n\geq n_{0}$, which implies that $\mathcal{H}^{s}(F)\leq \mathcal{H}_{n,3}^{s}(F)+\varepsilon$ for all $n\geq n_{0}$. Thus, by taking limits
we obtain the following expression:
\begin{equation}\label{eq:35}
\mathcal{H}^{s}(F)\leq \mathcal{H}_{3}^{s}(F)+\varepsilon
\end{equation}
On the other hand, the expression \ref{eq:34} also implies that $\mathcal{H}^{s}(F)\geq \mathcal{H}_{n}^{s}(F)-\varepsilon$ for all $n\geq n_{0}$. Therefore, an analogous argument to the previous one leads to
\begin{equation}\label{eq:36}
\mathcal{H}^{s}(F)\geq \mathcal{H}_{3}^{s}(F)-\varepsilon
\end{equation}
Finally, since the arbitrariness of $\varepsilon$ and by means of the expressions \ref{eq:35} and \ref{eq:36}, we obtain that $\mathcal{H}^{s}(F)=\mathcal{H}^{s}_{3}(F)$.
\end{proof}
Note that the theoretical method we have just introduced on definition \ref{eq:69} allows to calculate the fractal dimension (if exists) of a subset by taking into account the diameters of all the elements of each family $\mathcal{A}_{n}(F)$.
Accordingly, it provides a more accurate model in order to determine the fractal dimension of a subset than by means of fractal dimension I or II models, where we only had into account the number of elements of each level of the fractal structure which meet the given subset.

Recall that the box-counting dimension can be estimated on an easy way which allows its effective implementation on any programming language. Thus, the fractal dimension I \& II models have inherited its advantage, which leads to some interesting applications where they can provide information that box-counting method cannot offer (see \cite[Section 5]{SGFD}).
Indeed, the regression line of $N_{n}(F)$ versus $n$ on a logarithmic scale gives the fractal dimension by means of its slope. Nevertheless, as it happens with the box-counting model, the fractal dimensions I \& II present no good analytical properties as Hausdorff dimension has.
Therefore, as well as Hausdorff dimension is based on its corresponding measure $\mathcal{H}_{H}$, it results also interesting to determine some analytical properties for the set functions $\mathcal{H}^{s}$ and $\mathcal{H}_{3}^{s}$, which leads to fractal dimension III.

First of all, denote by $\mathcal{P}(X)$ to the class of all subsets of a given space $X$.
Recall that an outer measure $\mu$ is a set function from $\mathcal{P}(X)$ to $[0,\infty]$ which verifies the next properties: (i) it assigns the value $0$ to the empty set, (ii) it is a monotonic increasing mapping, and (iii) it is countably additive, namely, it verifies that $\mu(\cup_{n\in \mathbb{N}}A_{n})\leq \sum_{n=1}^{\infty}\mu(A_{n})$ for all sequence $\{A_{n}\}_{n\in \mathbb{N}}$ of subsets of $X$ .
Thus, it can be checked that $\mathcal{H}_{n}^{s}$ is an outer measure for all $n\in \mathbb{N}$. In order to see that $\mathcal{H}_{n,3}^{s}$ also is, it results useful \cite[Theorem 5.2.2]{EDG90}, since it suffices with taking $\mathcal{A}=\cup_{m\geq n}\{\mathcal{A}_{m}(F)\}$, $\textbf{c}:\mathcal{A}\rightarrow [0,\infty]$ as the set function
which maps $A$ to $\delta(A)^{s}$ for all $A\in \mathcal{A}_{n,3}(F)$, and define $\overline{\mu}:\mathcal{P}(X)\rightarrow [0,\infty]$ in the same way as $\mathcal{H}_{n,3}^{s}$ for all $n\in \mathbb{N}$. Nevertheless, although these two set functions are outer measures, it is not true in general that their limits also are, as the next remark establishes.
\begin{obs}
Neither $\mathcal{H}^{s}$ nor $\mathcal{H}_{3}^{s}$ are outer measures.
\end{obs}

\begin{proof}
Indeed, let $\ef$ be the natural fractal structure on the real line induced on $[0,1]$, namely, $\ef=\{\Gamma_{n}:n\in \mathbb{N}\}$ where $\Gamma_{n}=\{[\frac{k}{2^{n}},\frac{k+1}{2^{n}}]:k\in \{0,1,\ldots ,2^{n}-1\}\}$ for all natural number $n$, and consider $F=\mathbb{Q}\cap [0,1]$. It is clear that $\mathcal{H}_{n}^{s}(\{q_{i}\})\leq \frac{1}{2^{ns-1}}$ for all $q_{i}\in F$ and all $n\in \mathbb{N}$.
On the other hand, since each element of the level $\Gamma_{n}$ of the fractal structure contains a rational $q\in F$, then we have that $\mathcal{H}_{n}^{s}(F)=\frac{1}{2^{n(s-1)}}$ for all $n\in \mathbb{N}$. In this way, for all $s\in (0,1)$, we have that $\mathcal{H}^{s}(\{q_{i}\})=0$ for any $q_{i}\in F$, as well as we get that $\mathcal{H}^{s}(F)=\infty$. Accordingly, it is not verified the countable additive property.
Note that the same counterexample remains valid in order to justify that the set function $\mathcal{H}_{3}^{s}$ is not also an outer measure.
\end{proof}

In this way, an interesting question which appears on a natural way, consists of determining some reasonable properties over the elements of the fractal structure, in order to relate the new definition of fractal dimension with those studied in \cite{SGFD}.
Taking it into account, the next theorem results useful in that direction, since it ensures that fractal dimension III is going to agree with fractal dimension I when working with fractal structures whose elements have an appropiate diameter.

In order to show it, recall that two sequences of positive real numbers $\{f(n)\}_{n\in \mathbb{N}}$ and $\{g(n)\}_{n\in \mathbb{N}}$ have the same order, which we denote by $\mathcal{O}(f)=\mathcal{O}(g)$, if and only if $\lim_{n\rightarrow \infty}\frac{f(n)}{g(n)}\in (0,\infty)$.
The following technical lemma, whose straightforward proof is left to the reader, is going to be useful for our purposes.

\begin{lema}\label{prop:4}
Let $f,g:\mathbb{N}\rightarrow \mathbb{R}^{+}$ be two sequences of positive real numbers such that $\mathcal{O}(f)=\mathcal{O}(g)$, and suppose that there exists $\lim_{n\rightarrow \infty}\frac{h(n)}{f(n)}\in (0,\infty)$ with $h:\mathbb{N}\rightarrow \mathbb{R}^{+}$. Then there exists a constant $k\in (0,\infty)$ such that
\begin{equation}
\lim_{n\rightarrow \infty}\frac{h(n)}{f(n)}=k\cdot \lim_{n\rightarrow \infty}\frac{h(n)}{g(n)}
\end{equation}
\end{lema}


\begin{teo}\label{teo:5}
Let $\ef$ be a fractal structure on a metric space $(X,\rho)$ and let $F$ be a subset of $X$. Suppose that $\delta(A)=\delta(F,\Gamma_{n})$ for all $A\in \mathcal{A}_{n}(F)$ and $\mathcal{O}(\delta(F,\Gamma_{n}))=\mathcal{O}(\frac{1}{2^{n}})$ for all $n\in \mathbb{N}$. If the fractal dimension I of $F$ exists, then $\dim_{\ef}^{1}(F)=\dim_{\ef}^{3}(F)$.
\end{teo}

\begin{proof}
First of all, let $\varepsilon$ be a fixed but arbitrary positive real number, and denote $\alpha=\dim_{\ef}^{1}(F)$. Thus, there exists a natural number $n_{0}$ such that
\begin{equation}\label{eq:72}
2^{n(\alpha-\varepsilon)}\leq N_{n}(F)\leq 2^{n(\alpha+\varepsilon)}
\end{equation}
for all $n\geq n_{0}$. Moreover, note that
\begin{equation}\label{eq:73}
\mathcal{H}_{n,3}^{s}(F)
\leq N_{m}(F)\cdot \delta(F,\Gamma_{m})^{s}
\leq 2^{m(\alpha+\varepsilon)}\cdot \delta(F,\Gamma_{m})^{s}
\end{equation}
for $m\geq n\geq n_{0}$, by means of expression \ref{eq:72}.
Hence, it suffices with taking limits as $n\rightarrow \infty$ at \ref{eq:73} in order to get what follows:
\begin{equation}\label{eq:38}
\begin{split}
\mathcal{H}^{s}_{3}(F)
&=\lim_{n\rightarrow \infty} \mathcal{H}_{n,3}^{s}(F)\\
&\leq \lim_{m\rightarrow \infty} \Big(2^{m(\alpha+\varepsilon)}\cdot \delta(F,\Gamma_{m})^{s}\Big)\\
&=k^{s}\cdot \lim_{m\rightarrow \infty}2^{m(\alpha+\varepsilon-s)}\\
&=\left\{
\begin{array}{ll}
\hbox{$\infty$}\ \ldots & \hbox{$\alpha+\varepsilon> s$}\\
\hbox{$0$}\ \ldots & \hbox{$\alpha+\varepsilon< s$}
\end{array}
\right.
\end{split}
\end{equation}
where we have used lemma \ref{prop:4} in order to get the second equality on \ref{eq:38}.
Accordingly, it results clear that
\begin{equation}\label{eq:31}
\dim_{\ef}^{3}(F)\leq \dim_{\ef}^{1}(F)+\varepsilon
\end{equation}
On the other hand, given a fixed but arbitrary real number $\delta>0$, we have that for all $n\in \mathbb{N}$ there exists a natural number $m(n)\geq n$ such that
\begin{equation}\label{eq:74}
\mathcal{H}_{n,3}^{s}(F)+\delta
\geq N_{m(n)}(F)\cdot \delta(F,\Gamma_{m(n)})^{s}
\geq 2^{m(n)(\alpha-\varepsilon)}\cdot \delta(F,\Gamma_{m(n)})^{s}
\end{equation}
where we have applied \ref{eq:72} in order to get the second inequality at \ref{eq:74}. Therefore, by taking limits as $n\rightarrow \infty$ on the previous expression, we get:
\begin{equation}\label{eq:75}
\begin{split}
\mathcal{H}^{s}_{3}(F)+\delta
&\geq \lim_{m(n)\rightarrow \infty}\Big(N_{m(n)}(F)\cdot \delta(F,\Gamma_{m(n)})^{s}\Big)\\
&\geq \lim_{m(n)\rightarrow \infty} \Big(2^{m(n)(\alpha-\varepsilon)}\cdot \delta(F,\Gamma_{m(n)})^{s}\Big)\\
&=\rho^{s}\cdot \lim_{m(n)\rightarrow \infty}2^{m(n)((\alpha-\varepsilon)-s)}\\
&=\left\{
\begin{array}{ll}
\hbox{$\infty$}\ \ldots & \hbox{$\alpha-\varepsilon> s$}\\
\hbox{$0$}\ \ldots & \hbox{$\alpha-\varepsilon< s$}
\end{array}
\right.
\end{split}
\end{equation}
Note that lemma \ref{prop:4} has been used at the first equality of \ref{eq:75}.
Thus, if we consider $\alpha-\varepsilon>s$, then we have that $\mathcal{H}_{3}^{s}(F)+\delta=\infty$, and by means of the arbitrariness of $\delta$, we have that $\mathcal{H}_{3}^{s}(F)=\infty$. Accordingly, we obtain the following expression:
\begin{equation}\label{eq:32}
\dim_{\ef}^{1}(F)-\varepsilon\leq \dim_{\ef}^{3}(F)
\end{equation}
Finally, by means of \ref{eq:31} and \ref{eq:32} and taking into account the arbitariness of $\varepsilon>0$, we get the fact that fractal dimensions I \& III agree.
\end{proof}

\begin{obs}\label{obs:5}
Note that whenever fractal dimension I of $F$ does not exist, then theorem \ref{teo:5} asserts that $\underline{\dim}_{\ef}^{1}(F)=\dim_{\ef}^{3}(F)$ when working with fractal structures $\ef$ whose elements have a diameter of order $\frac{1}{2^{n}}$ on each level of $\ef$. Indeed, it suffices with taking into account the fact that if $\alpha=\underline{\dim}_{\ef}^{1}(F)$, then there exists a subsequence $\{\frac{\log N_{n_{k}}(F)}{n_{k}\log 2}\}_{n_{k}\in \mathbb{N}}$ of $\{\frac{\log N_{n}(F)}{n \log 2}\}_{n\in \mathbb{N}}$ such that $\alpha=\lim_{k\rightarrow \infty}\frac{\log N_{n_{k}}(F)}{n_{k}\log 2}$. Thus, by means of a similar argument to that used on the proof of theorem \ref{teo:5}, we get the result.
\end{obs}

We also suggest another condition over the size of the elements of each level of the fractal structure in order to get that fractal dimensions II \& III agree.
The proof of this result may be dealt with in a similar way than the previous one.
\begin{teo}\label{teo:6}
Let $\ef$ be a fractal structure on a metric space $(X,\rho)$, and let $F$ be a subset of $X$. Suppose that $\delta(F,\Gamma_{n})\rightarrow 0$ and there exists a natural number $n_{0}$ such that $\delta(A)=\delta(F,\Gamma_{n})$ for all $A\in \mathcal{A}_{n}(F)$ and all $n\geq n_{0}$. If the fractal dimension II of $F$ exists, then $\dim_{\ef}^{2}(F)=\dim_{\ef}^{3}(F)$.
\end{teo}

\begin{obs}\label{obs:6}
As seen on remark \ref{obs:5} for theorem \ref{teo:5}, we have now that whenever fractal dimension II of $F$ does not exist, then theorem \ref{teo:6} leads to $\underline{\dim}_{\ef}^{2}(F)=\dim_{\ef}^{3}(F)$. 
\end{obs}

Moreover, as an immediate consequence of remarks \ref{obs:5} and \ref{obs:6}, we can show the fact that fractal dimension III generalizes fractal dimensions I \& II over GF-spaces whose elements have order equal to $\frac{1}{2^{n}}$ on each level of the fractal structure we are working with.
\begin{cor}
Let $\ef$ be a fractal structure on a metric space $(X,\rho)$, with $F$ being a subset of $X$, and suppose that $\delta(A)=\delta(F,\Gamma_{n})$ for all $A\in \mathcal{A}_{n}(F)$ and $\mathcal{O}(\delta(F,\Gamma_{n}))=\mathcal{O}(\frac{1}{2^{n}})$ for all $n\in \mathbb{N}$. Then, $\underline{\dim}_{\ef}^{1}(F)=\underline{\dim}_{\ef}^{2}(F)=\dim_{\ef}^{3}(F)$.
\end{cor}


Nevertheless, note that although the fractal dimension III definition is inspired on a suitable \emph{discretization} of the model of Hausdorff dimension, it is going to agree with the box-counting dimension on the context of any euclidean space equipped with its natural fractal structure, as the following theorem stablishes.
\begin{teo}\label{teo:8}
Let $\ef$ be the natural fractal structure on the euclidean space $\mathbb{R}^{d}$, and let $F$ be a subset of $\mathbb{R}^{d}$. Then
\begin{equation*}
\underline{\dim}_{\ef}^{1}(F)=\underline{\dim}_{\ef}^{2}(F)=\dim_{\ef}^{3}(F)=\underline{\dim}_{B}(F)
\end{equation*}
\end{teo}
\begin{proof}
First of all, \cite[Theorem 4.5]{SGFD} leads to $\underline{\dim}_{\ef}^{1}(F)=\underline{\dim}_{\ef}^{2}(F)=\underline{\dim}_{B}(F)$. On the other hand, remark \ref{obs:5} asserts that $\underline{\dim}_{\ef}^{1}(F)=\dim_{\ef}^{3}(F)$ since all the elements of the level $\Gamma_{n}$ of the natural fractal structure on $\mathbb{R}^{d}$ have a diameter equal to $\frac{\sqrt{d}}{2^{n}}$.
\end{proof}
Accordingly, by means of theorem \ref{teo:8} we have that fractal dimension III generalizes both box-counting and fractal dimensions I \& II on the euclidean spaces. In this way, note that this result allows to calculate the fractal dimension III of a subset of $\mathbb{R}^{d}$ by means of the easier formulas applied for calculating both box-counting and fractal dimensions I \& II.
Moreover, in parallel with \cite[Proposition 3.4]{SGFD} for fractal dimension I (and also for fractal dimension II) the following properties are verified for the new fractal dimension defined on \ref{eq:69}.
In this way, note that theorem \ref{teo:8} provides an interesting tool in order to find suitable theoretical counterexamples.

\begin{prop}
Let $\ef$ be a fractal structure on a metric space $(X,\rho)$ and suppose that $\delta(F,\Gamma_{n})\rightarrow 0$ for all subset $F$ of $X$. Then,

\begin{enumerate}
\item $\dim_{\ef}^{3}$ is monotonic.
\item There exists countable sets $F\subset X$ such that $\dim_{\ef}^{3}(F)\neq
0$.
\item $\dim_{\ef}^{3}$ is not countably stable.
\item There exists a locally finite tiling starbase fractal structure $\ef$ with \label{it:5}
finite order on a suitable space $X$ such that
$\dim_{\ef}^{3}(F)\neq \dim_{\ef}^{3}(\overline{F})$ for a given subset
$F\subset X$.
\end{enumerate}
\end{prop}

Recall that on \cite[theorem 4.9]{SGFD} we have found some interesting properties over the elements of each level of a fractal structure in order to get the equality between fractal dimension II and box-counting dimension on a generic GF-space, taking into account the fact that the latter can be defined on any metrizable space.
In other words, for fractal dimension III the generalization of theorem \ref{teo:8} to the context of generic GF-spaces results interesting.
In order to do this, recall that there exists a useful property on each euclidean space which can be transfered to the context of fractal structures: given a positive real number $\delta$ and a subset $F$ of an euclidean space $\mathbb{R}^{d}$ verifying that $\delta(F)\leq \delta$, then it is verified that the number of $\delta-$cubes on $\mathbb{R}^{d}$ which meet $F$ is as most $3^{d}$.
\begin{teo}\label{cor:11}
Let $\ef$ be a fractal structure on a metric space $(X,\rho)$ and let $F$ be a subset of $X$. Suppose that there exists a natural number $k$ such that for all $n\in \mathbb{N}$, every subset $A$ of $X$ with $\delta(A)\leq \delta(F,\Gamma_{n})$ meets at most $k$ elements of each level $\Gamma_{n}$ of the fractal structure $\ef$. Suppose also that $\delta(F,\Gamma_{n})\rightarrow 0$.
Then, if $\delta(A)=\delta(F,\Gamma_{n})$ for all $A\in \mathcal{A}_{n}(F)$, we have that $\dim_{B}(F)=\dim_{\ef}^{3}(F)$.
\end{teo}

\begin{proof}
Note that theorem \ref{teo:6} leads to $\dim_{\ef}^{2}(F)=\dim_{\ef}^{3}(F)$. On the other hand, \cite[theorem 4.9]{SGFD} completes the proof.
\end{proof}

Another interesting question which have been studied by applying box-counting dimension (see \cite{FAL90}) and by means of fractal dimensions I \& II (see \cite{SGFD}) is about the possibility of computing the fractal dimension of a strict self-similar set by means of an easy expression.
The following restriction over an IFS has been widely used on fractal dimension theory (see \cite{FAL90} and \cite{SCH94}).

\begin{defn}
Let $I=\{1,\ldots, m\}$ be a finite index set with $(X,\{f_{i}:i\in I\})$ being an IFS whose associated self-similar set is $K$. It is said that the contractions $f_{i}$ satisfy the open set condition iff there exists a non-empty bounded open set $V$ of $X$ such that $\cup_{i\in I}f_{i}(V)\subset V$, with $f_{i}(V)\cap f_{j}(V)=\emptyset$ for all $i\neq j$.
\end{defn}

The next theorem becomes the classical one studied in \cite[Theorem 9.3]{FAL90}.

\begin{teo}\label{teo:9}
Let $I=\{1,\ldots, m\}$ be a finite index set, with $(\mathbb{R}^{d},\{f_{i}:i\in I\})$ being an IFS whose associated strict self-similar set is $K$, and let $c_{i}$ be the contraction factors associated with the similarities $f_{i}$ which satisfy the open set condition.
Then $\dim_{B}(F)=\dim_{H}(F)=s$, where $s$ is the solution of the following equation
\begin{equation}\label{eq:77}
\sum_{i\in I}c_{i}^{s}=1
\end{equation}
Furthermore, $s$ verifies that $\mathcal{H}_{H}^{s}(F)\in (0,\infty)$.
\end{teo}
The open set condition is an important restriction which we require to the pieces of a self-similar set in order to ensure that they \emph{do not overlap too much}.
Indeed, when this property is satisfied for the similarities of the corresponding IFS, then a parallel result to theorem \ref{teo:9} can be shown for fractal dimension II definition (see \cite[Theorem 4.15]{SGFD}).
However, by means of fractal dimension III it is also possible to be able to calculate the fractal dimension of a strict self-similar set by means of an easy expression like \ref{eq:77}, even if its similarities does not satisfy the open set condition, which allows to generalize theorem \ref{teo:9} to the context of GF-spaces.
The proof of such result is based on the natural fractal structure associated with a self-similar set and also takes into account the definition of $\mathcal{H}_{3}^{s}$.

\begin{teo}\label{teo:7}
Let $I=\{1,\ldots, k\}$ be a finite index set, $X$ be a complete metric space and $(X,\{f_{i}:i\in I\})$ be an IFS whose associated strict self-similar set is $K$. Suppose that $c_{i}$ are the similarity factors associated with the similarities $f_{i}$, and let $\ef$ be the natural fractal structure on $K$ as a self-similar set.
Then, $\dim_{\ef}^{3}(K)=s$ where $s$ is given by
$\sum_{i\in I}c_{i}^{s}=1$,
and for this value of $s$ we have also that $\mathcal{H}^{s}_{3}(K)\in (0,\infty)$.
\end{teo}
\begin{proof}
First of all, since $K$ is the self-similar set given by the finite set of similarities $\{f_{i}:i\in I\}$, we have that it is the unique non-empty compact subset of $X$ which verifies the next Hutchinson's equation:
$K=\cup_{i\in I}f_{i}(K)$. Note
that this argument also implies that
$\mathcal{A}_{n,3}(K)=\cup_{m\geq n}\{\Gamma_{m}\}$
for all $n\in \mathbb{N}$.
On the other hand, let $s$ be a non-negative real number such that $\sum_{i\in I}c_{i}^{s}=1$, and denote by $J_{l}$ to the set $\{(i_{1},\ldots, i_{l}):i_{j}\in I, 1\leq j\leq l\}$.
Accordingly, if we denote $K_{i_{1}\ldots i_{l}}=f_{i_{1}}\ldots f_{i_{l}}(K)$, then we get $K=\cup_{J_{l}}K_{i_{1}\ldots i_{l}}$.
Moreover, since $c_{i}$ is the similarity factor associated with each similarity $f_{i}$, then $c_{i_{1}}\ldots c_{i_{l}}$ is the similarity factor associated with the composition of similarities $f_{i_{1}}\circ f_{i_{2}}\circ \ldots \circ f_{i_{l}}$. Taking it into account, we get
\begin{equation}
\begin{split}
\mathcal{H}_{n,3}^{s}(K)
&=\inf\Bigg\{\sum_{A\in \mathcal{A}_{m}(K)}\delta(A)^{s}:m\geq n\Bigg\}\\
&=\inf\Bigg\{\sum_{(i_{1},\ldots, i_{m})\in J_{m}}\delta(K_{i_{1}\ldots i_{m}})^{s}:m\geq n\Bigg\}\\
&=\inf\Bigg\{\Bigg(\sum_{i_{1}\in I}c_{i_{1}}^{s}\Bigg)\ldots \Bigg(\sum_{i_{m}\in I}c_{i_{m}}^{s}\Bigg) \ \delta(K)^{s}:m\geq n\Bigg\}\\
&=\delta(K)^{s}
\end{split}
\end{equation}
for all natural number $n$, which implies that $\mathcal{H}_{3}^{s}(K)=\delta(K)^{s}$. Thus, $\mathcal{H}_{3}^{s}(K)\notin \{0,\infty\}$, namely, $s$ is the point where $\mathcal{H}_{3}^{s}(K)$ \emph{jumps} from $\infty$ to $0$, so that $s=\dim_{\ef}^{3}(K)$.
\end{proof}

Moreover, we affirm that the hypothesis consisting of the fact that $K$ is a strict self-similar set results necessary as the next counterexample shows.

\begin{obs}\label{obs:3}
Let $I=\{1,\ldots,8\}$ be a finite index set and let $(\mathbb{R}^{2},\{f_{i}:i\in I\})$ be an IFS whose associated self-similar set is $K=[0,1]^{2}$. Consider the contractions $f_{i}:\mathbb{R}^{2}\rightarrow \mathbb{R}^{2}$ defined by
\begin{equation}
f_{i}(x,y)=\left\{
\begin{array}{ll}
\hbox{$(\frac{-y}{2},\frac{x}{4})+(\frac{1}{2},\frac{i-1}{4})$}\ \ldots & \hbox{$i=1,2,3,4$} \\
\hbox{$(\frac{-y}{2},\frac{x}{4})+(1,\frac{i-5}{4})$}\ \ldots & \hbox{$i=5,6,7,8$}
\end{array}
\right.
\end{equation}
Let also $\ef$ be the natural fractal structure on $K$ as a self-similar set. Then, $\dim_{\ef}^{3}(K)\neq s$ where $s$ is the solution of the equation $\sum_{i\in I}c_{i}^{s}=1$.
\end{obs}

\begin{proof}
First of all, note that the self-similar set $K$ is not a strict one.
Furthermore, the contractions $f_{i}$ are compositions of affinity mappings: rotations, dilations (in the plane and respect to one coordinate) and translations. In this way, it is also clear that all the contractive mappings $f_{i}$ have the same contraction factor, equal to $\frac{1}{2}$.
Accordingly, the solution of the equation $\sum_{i\in I}c_{i}^{s}=1$ leads to $s=3$.\newline
On the other hand, we affirm that $\dim_{\ef}^{3}(K)=2$.
In order to show it, we are going to calculate the fractal dimension III of $K$ by means of theorem \ref{cor:11}. Indeed, we begin by taking into account all the even levels of the fractal structure $\ef$. In this way, note that for all $n\in \mathbb{N}$, every level $\Gamma_{2n}$ is composed by squares whose sides are equal to $\frac{1}{8^{n}}$. Moreover, it is also clear that $\delta(A)=\delta(K,\Gamma_{2n})=\frac{\sqrt{2}}{8^{n}}$ for all $A\in \Gamma_{2n}$, which implies that $\delta(K,\Gamma_{2n})\rightarrow 0$.
Now, we have to check the main condition on theorem \ref{cor:11}.
In this way, it suffices with calculating the maximum number of elements of $\Gamma_{2n}$ which are intersected by a subset $B$ whose diameter is at most equal to $\frac{\sqrt{2}}{8^{n}}$ for all natural number $n$.
Indeed, the relation between the diameter of each square of any level $\Gamma_{2n}$ of the fractal structure and its side leads to $\frac{\sqrt{2}\cdot \frac{1}{8^{n}}}{\frac{1}{8^{n}}}=\sqrt{2}<2$, which implies that the number of elements of $\mathcal{A}_{2n}(B)$ is at most $3$ on each direction for all subset $B$ with $\delta(B)\leq \delta(K,\Gamma_{2n})$. Accordingly, we can choose $k=3\cdot 3=9$ as a suitable constant for the levels of even order on the fractal structure we are working with.\newline
Similarly, it can be checked that all the levels of odd order $\Gamma_{2n+1}$ of the fractal structure $\ef$ are composed by rectangles whose dimensions are $\frac{1}{2\cdot 8^{n}}\times \frac{1}{4\cdot 8^{n}}$ for all $n\in \mathbb{N}$. Note also that all the elements on each level $\Gamma_{2n+1}$ have the same diameter, equal to $\frac{\sqrt{5}}{4\cdot 8^{n}}$, so that the sequence of diameters $\delta(K,\Gamma_{2n+1})$ also converges to $0$.
In order to check the last condition, we have the next relations between diameter and sides of each rectangle: on the one hand, $\frac{\frac{1}{4}\cdot \frac{\sqrt{5}}{8^{n}}}{\frac{1}{2}\cdot \frac{1}{8^{n}}}=\frac{\sqrt{5}}{2}<2$, and on the other hand, $\frac{\frac{1}{4}\cdot \frac{\sqrt{5}}{8^{n}}}{\frac{1}{4}\cdot \frac{1}{8^{n}}}=\sqrt{5}<3$. Therefore, each subset $A$ whose diameter is at most equal to $\delta(K,\Gamma_{2n+1})$ has to meet at most at $k=3\cdot 4=12$ elements of each level $\Gamma_{2n+1}$.\newline
Accordingly, the main hypothesis on the theorem \ref{cor:11} is also satisfied, since it suffices with taking $k=12$ as a suitable constant for any level of the fractal structure. Thus, we have that $\dim_{B}(K)=\dim_{\ef}^{3}(K)=2$.
\end{proof}

Furthermore, by means of \cite[theorem 9.3]{FAL90} as well as theorem \ref{teo:7}, we get an interesting result whose proof becomes now immediate: both box-counting and Hausdorff dimensions agree with fractal dimension III if the similarities associated with the corresponding IFS verify the open set condition.

\begin{cor}\label{cor:9}
Let $I=\{1,\ldots, m\}$ be a finite index set and let $(\mathbb{R}^{d},\{f_{i}:i\in I\})$ be an IFS whose associated strict self-similar set is $K$. Suppose that $c_{i}$ are the similarity factors associated with the similarities $f_{i}$ which satisfy the open set condition, and let $\ef$ be the natural fractal structure on the self-similar set $K$.
Then, $\dim_{\ef}^{3}(K)=\dim_{B}(K)=\dim_{H}(K)$.
\end{cor}

Next, by means of a suitable counterexample we show that corollary \ref{cor:9} is not improvable in the sense that we cannot remove the hypothesis consisting of the open set condition. Indeed,

\begin{obs}\label{obs:7}
There exists a strict self-similar set $K$ provided with its natural fractal structure, whose similarities $f_{i}$ do not satisfy the open set condition, and whose Hausdorff dimension and fractal dimension III do not agree.
\end{obs}

\begin{proof}
Let $I=\{1,2,3\}$ be a finite index set and let $(\mathbb{R},\{f_{i}:i\in I\})$ be an IFS whose associated self-similar set $K$ is the closed unit interval on the real line, which verifies the next Hutchinson's equation:
\begin{equation}
K=f_{1}(K)\cup f_{2}(K)\cup f_{3}(K)
\end{equation}
where $f_{i}:\mathbb{R}\rightarrow \mathbb{R}$ are the contractive mappings given by
\begin{equation}
f_{i}(x)=\left\{
\begin{array}{ll}
\hbox{$\frac{x}{2}$}\ \ldots & \hbox{$i=1$} \\
\hbox{$\frac{x+1}{2}$}\ \ldots & \hbox{$i=2$} \\
\hbox{$\frac{2x+1}{4}$}\ \ldots & \hbox{$i=3$}
\end{array}
\right.
\end{equation}
Let also $\ef$ be the natural fractal structure on $K$ as a self-similar set.
First of all, note that all the contractions $f_{i}$ are similarities with all the similarity factors equal to $\frac{1}{2}$. Thus, $K$ is a strict self-similar set on $\mathbb{R}$.
Moreover, by means of theorem \ref{teo:7} we have that $\dim_{\ef}^{3}(K)$ is the solution of the equation $\sum_{i\in I}c_{i}^{s}=1$, which leads to $\dim_{\ef}^{3}(K)=\frac{\log 3}{\log 2}$.
Also, note that theorems \ref{teo:5} and \cite[Theorem 4.11]{SGFD} imply that $\dim_{\ef}^{1}(K)=\dim_{\ef}^{2}(K)=\frac{\log 3}{\log 2}=\dim_{\ef}^{3}(K)$ since all the elements on $\ef$ have a diameter whose order is $\frac{1}{2^{n}}$ on each level of the fractal structure.
Finally, we affirm that the similarities $\{f_{i}:i\in I\}$ does not satisfy the open set condition. Indeed, suppose the opposite. Then, by means of corollary \ref{cor:9}, we have that $\dim_{\ef}^{3}(K)$ is equal to $\dim_{H}(K)$ which is a contradiction.
\end{proof}

Remark \ref{obs:7} provides a suitable example of a subset of a GF-space whose fractal dimension III and Hausdorff dimension do not agree.
In this way, we also know the fact that fractal dimension III can be different from fractal dimensions I \& II as the following remark establishes.
\begin{obs}\label{obs:8}
There exists a strict self-similar set $K$ provided with its natural fractal structure, such that its fractal dimensions I \& II are different from its fractal dimension III.
\end{obs}

\begin{proof}
Let $I=\{1,2,3\}$ be a finite index set and let $(\mathbb{R},\{f_{i}:i\in I\})$ be an IFS whose associated self-similar set $K$ is the closed unit interval on the real line, which satisfies the next Hutchison's equality:
\begin{equation*}
K=\cup_{i\in I}f_{i}(K)
\end{equation*}
where $f_{i}:\mathbb{R}\rightarrow \mathbb{R}$ are the contractions given by
\begin{equation*}
f_{i}(x)=\left\{
\begin{array}{ll}
\hbox{$\frac{x}{2}$}\ \ldots & \hbox{$i=1$} \\
\hbox{$\frac{x+2}{4}$}\ \ldots & \hbox{$i=2$} \\
\hbox{$\frac{x+3}{4}$}\ \ldots & \hbox{$i=3$}
\end{array}
\right.
\end{equation*}
Let also $\ef$ be the natural fractal structure on $K$ as a self-similar set. It is clear that the  mappings $f_{i}$ are similarities, so $K$ is a strict self-similar set. On the one hand, since the open set condition is verified (it suffices with taking $V=(0,1)$ as a suitable open set), an application of the corollary \ref{cor:9} leads to $\dim_{\ef}^{3}(K)=\dim_{B}(K)=\dim_{H}(K)=1$ (since $K=[0,1]$). On the other hand, note that each level $\Gamma_{n}$ of the fractal structure contains $3^{n}$ subintervals of $[0,1]$. Taking also into account that $\delta(K,\Gamma_{n})=\frac{1}{2^{n}}$ for all $n\in \mathbb{N}$, we have that $\dim_{\ef}^{2}(K)=\frac{\log 3}{\log 2}=\dim_{\ef}^{1}(K)$.
\end{proof}


\end{document}